\documentclass[11pt]{article}
\usepackage{cite}

\title{The Equation of Atom Motion  in an External Gravitational Field.}
\author{Basalyga G.V.{\thanks{E-mail: basalyga@hotmail.com}} , Gorbatsievich A.K.{\thanks{E-mail: gorbatsievich@phys.bsu.unibel.by }}
\\
\\ {Department of Theoretical Physics,
  Byelorussian State  University,}
\\ {220080 Minsk, Belarus }}

\date{January 28, 1999}

\begin {document}

\maketitle

\begin{abstract}
It  is  shown  that  the  motion    of    a
multielectron atom in an external  gravitational   field  in a good
approximation    is    described    by    system    of        the
Mathisson---Papapetrou  equations,  if  we  put  as  a  classical
angular momentum of the atom the expectation value of the  operator
of the full angular momentum of the system, which includes  spins
of the nucleus  and  electrons,  and  orbital  momentums  of  the
electrons in the atom.
\end{abstract}

\newpage
\section {Introduction}

The interest to the problem of the atom  motion  in
an external gravitational fields is caused by the  following  two
reasons.

First, the question about the possibility of using the  atoms  as
the detectors of the gravitational  fields  represents  the
significant interest  for  modern  astrophysics
\cite{dalgar, pino1, pino2, parker1}.
Second, because the atom is an extended  quantum  system,
with the inner structure
(the  motion  of
extended bodies in general relativity is widely discussed in  the
literature),  the  problem  of
 atom motion has considerable theoretical interest too.

The analysis of this situation shows, that the  influence  of  an
external  gravitational field on  the  atomic spectra   can
reach the experimentally  measurable  values  only  in  extremely
strong  gravitational  fields  (with  characteristic  radii    of
curvature  $\leq$$10^{-3}$  cm)  \cite{parker2,  parker3,
gorb}, or in the case of  ultrarelativistic  motion  of  atom  as
a whole \cite{gorb, gorb2}. Apparently, in both cases the  motion
of an atom cannot be considered {\em a priori}  as  a  geodesic
one. So the finding of the  equations  of  the  atom  motion  in  an
arbitrary curved space-time is the important  problem.
For the case of the one-electron atom this  problem  was
solved in the paper \cite {gorb1}. Following this work  we shall generalize
received results   for the  case  of  multielectron  atom.

For essential simplification of the subsequent  calculations,  we
shall use two  assumptions. First, because nucleus mass $\mu$ is
much greater than electron mass $m$ ($\mu$$\gg$$m$),
the atom motion can be considered in a good approximation as
a motion of a classical point particle  with  the  law  of
motion
\begin  {equation}
x^i=\xi ^i(\tau )
\label  {1}
\end{equation}
where $\tau$ is the proper  time:  $d\tau  ^2=-c^{-2}  ds^2 $.
Moreover, the  assumed  identity  of  atom  and  nucleus  masses
permits the consideration of the atomic motion as a nucleus one.
Second, we shall  describe the motion  of  electrons
in the atom  by  quantum  mechanics.  Moreover,
because the  relative  motion  of  electrons  in  the  atom  is
the essential non-relativistic ($v \ll c$), so in the special comoving
reference frame the  atom  can  be  described  in  frameworks  of
non-relativistic (or quasi-relativistic) quantum mechanics
\cite{gorb}. Subsequently we shall assume that the atom is in a 
quasistationary quantum state.

Thus, one can obtain the equation for $\xi^i(\tau  )$  from
the following suggested model.  The  nucleus,  in  the  classical
mass-point approximation with intrinsic angular momentum  (spin),
is moving in an external gravitational field and interacting with
the electromagnetic field of the electrons. The electrons, which for
theirs part is moving in external gravitational field and in  the
electromagnetic  field  of  the   nonexcited  and   electrons,    are
considered in a quasistationary bound quantum state.

 We notice, that  this  assumptions are  usual.  It  allows us
instead of the relativistic problem of many bodies, which even in
flat  space  has  a  lot  of  difficulties\footnote {Quantum
mechanical problem of two bodies in a  weak  gravitational  field
was discussed in work \cite{fish}.},  consider  the  motion  of
electrons in  the  given  electromagnetic  field,  created  by  a
nucleus, and also in an external gravitational field.

As it is shown in  works  \cite{audr,  gorb},  the  generalized
Ehrenfest  theorem  was  proved  for  a    Dirac    particle    in
quasi-classical approximation. According  to  this  theorem,  the
expectations values of the  position  and  spin  operators  of  a
electron  fulfill  the  Mathisson---Papapetrou   equations
\cite{mathis, papap},
which are describing a  motion  of  a  classical
particle with a spin in an external gravitational field. Applying
this results (valid for any Dirac particles) to  the  nucleus  is
represented justified. Therefore, to describe the nuclear  motion
in  our  model,  we  will  use  the system  of   Mathisson---Papapetrou
equations. For this purpose, in the first equation of this  system
we have  to  add  a  Lorentz  force,  which  describes  in  first
approximation the interaction of the  nucleus  with  the quantum state averaging
electromagnetic field of its  electrons.  Here,  we  neglect the
terms, which contain the second derivative of 4-velocity and describe interaction 
between the  nuclear  magnetic
moment and the  field  of  electrons  \cite  {dix},  because  the
numerical estimations show, that these terms are smaller than all
other terms of the equation. In result we receive  the  following
system of the  equations:
\begin  {equation}
\mu  \frac  {Du^i}{D\tau}  =\frac  1  {2c}  R^{i}\,_  {jkl}  u^j\varepsilon  ^   {klpq}
\bar{S} _pu_q + \frac {Ze} cF^ {ij} u_j,  \label  {2}
\end {equation}
\begin {equation}
a) \,  \frac {D\bar{S}  ^i}  {D\tau}
=\frac 1 {c^2} u^i\bar{S} _n\frac {Du^n} {D\tau},
\quad
b) \,    u^n\bar{S} _n=0, \label {3} \end {equation}
where
$Ze$, $u^i=\frac {dx^i} {d\tau} $ and $\bar{S}  _n$  denote
the charge, the 4-velocity and the classical spin of the  nucleus
respectively, here under the classical  spin  we  understand  the
expectation value of the nuclear spin  operator;  $\varepsilon  ^
{klpq}  $  is  the  Levi-Civita  pceudotensor  with  $\varepsilon
^{1234} = (-\det g_ {ij}) ^ {-1/2}$; $R^{i}\,_ {jkl}$ is the Riemann
curvature tensor;
 $F^{i}\,_{j} $  is the tensor of the electromagnetic field of electrons;
 $c$  is
the velocity of light. Here and everywhere the latin indexes run  from 1 to 4, and
greek ones from 1  to 3. The signature  of the space-time metric is $+2$.
We shall notice, that equation (\ref {3}b) represents the additional Pirani  condition
\cite{piran}
\footnote {By force of assumptions made above, the  
Pirani  condition  coincides with the Tulczyjew-Dixon condition \cite{dix, tulcz}.}. 
Due to this condition it is possible to enter
a 4-vector of a spin $\bar{S}_k$ and constant $\mu$,
which we shall identify hereinafter with nuclear mass.

  In the equation (\ref {2}) of  nuclear motion it  is unknown only one quantity
--  the tensor of the average electromagnetic field $F^ {ij} $, created by
nuclear electrons. Thus, the problem about finding of the equations of  motion
of atom in an external gravitational field is reduced to calculation of
 the electromagnetic field of nuclear electrons. The most simple
this problem can be decided in the special, comoving atom reference frame.

\section {The Reference Frame of the Single Observer.}

As a comoving reference frame we shall choose a reference frame of the
single observer \cite {sing}. From our point of view, this
reference frame is the most convenient for the description of objects of the
small sizes.

The reference frame of the single observer is defined by a motion of a single
mass point  (``the single observer''). The  world line of this mass point  is
named basis line. Along this line we establish a comoving tirade $h_ {(k)} ^i
(\tau) $, determined by the condition  $h_ {(4)} ^i=\frac 1cu^i$
accurate to three-dimensional rotations. If we will use the  world line of the
nucleus (\ref {1}) as a basis  line, we will obtain a comoving reference frame
for the atom.

Three-dimensional physical space in the reference frame of the single observer
 we shall define as a
geodesic spacelike hypersurface $f$, which is orthogonal to the basis line in
the given moment of proper time $\tau$. At each point P, laying on a
hypersurface $P\in f (\tau) $, we shall  put in conformity three scalars $X^
{(\alpha)} =\sigma _Ph_i^ {(\alpha)} k^i$, which will hereinafter play a role
of the three-dimensional coordinates. Here $\sigma_P$ is the value of the conical  parameter
$\sigma$ at the point P, defined along a spacelike geodesic in  $f$; $k^i$ is
the tangent unit vector to that geodesic at the point $\sigma=0$.

For nonrotating frame (i.e. when the tirade $h_ {(k)} ^i (\tau) $ is set along
a basic line with the help of Fermi---Walker transport) the quantities $ (X^
{(\alpha)}, c\tau) $ correspond to the Fermi normal coordinates, defined by
Synge \cite {sing, mizn}. In general case of rotating  reference frame,
the coordinates defined in the same way
\begin {equation}
x^ {\bar {\alpha}} =X^ {(\alpha)}, \quad
 x^ {\bar {4}} =c\tau,
\label {4} \end {equation}
we shall name as  rotating Fermi coordinates \cite {ni}. In these coordinates
the metric tensor can be presented as: \begin {equation}
g_ {\bar {i} \bar {j}} =\eta _ {(i) (j)} + \varepsilon _ {(i) (j)},
\label {5} \end {equation}
where
\begin {equation}
 \varepsilon _ {(\alpha) (\beta)} =-\frac {1}{3}R_ {(\alpha) (\mu)
(\beta) (\nu)} X^ {(\mu)} X^ {(\nu)} + O (\rho ^3), \label {6} \end {equation}
\begin {equation} \varepsilon _ {(\alpha) (4)} =\frac{1}{c}e_ {(\alpha) (\sigma)
(\tau)} X^ {(\tau)} \omega^ {(\sigma)} + \theta _ {(\alpha)}, \label {7}
\end{equation}
\begin {equation} \varepsilon _ {(4) (4)} =-2\left (\frac 1 {c^2} W_
{(\alpha)} X^ {(\alpha)} + \theta \right), \label {8} \end {equation}
\begin{equation}
\theta _ {(\alpha)} =\frac 23R_ {(\alpha) (\mu) (\nu) (4)} X^
{(\mu)} X^ {(\nu)} + O (\rho^3), \label {9} \end {equation}
 \begin {equation}
\theta = \frac{1}{2}\left (R_ {(4) (\mu) (4) (\nu)} -\frac 12R_ {(4) (\mu) (4)
(\nu) (\tau)} X^ {(\tau)} \right) X^ {(\mu)} X^ {(\nu)} + O (\rho ^4).
\label {10} \end {equation}
Here, we used the following notations: $\eta {(i) (j)} =diag (1,1,1,-1) $
is  the Minkowski tensor;
\begin{equation}
W^ {(\alpha)} =h_i^ {(\alpha)} \frac {Du^i} {D\tau}
\label {10a} \end {equation}
and
\begin{equation}
\omega ^ {(\alpha)} =\frac 12e^ {(\alpha) (\kappa) (\tau)} h_ {(\tau) i}
\frac {Dh_ {(\kappa)} ^i} {D\tau}
\label {10b} \end {equation}
  are the acceleration and the angular
velocity of the reference frame respectively;
$e^ {(\alpha) (\kappa) (\tau)} $ is the three-dimensional Levi-Civita
symbol;
$R^ {(i)}\,_ {(m) (n) (k)}$ and
\begin{equation}
R^ {(i)}\,_ {(m) (n) (k) (l)}=h_a^ {(i)} h_
{(m)} ^bh_ {(n)} ^ch_ {(k)} ^dh_ {(l)} ^pR^{a}\,_ {bcd; p}
\label {10c} \end {equation}
 are the  tirade
components of the Riemann  curvature  tensor  and its covariant derivative
respectively, determined along the basic line;
$\rho = \sqrt {X_ {(\alpha)} X^ {(\alpha)}} $. 

Because of the small sizes of atom, the condition
$\varepsilon _ {(i) (j)}\ll 1$ is  allowable even for  very strong (from the macroscopic point of view)
gravitational fields, therefore for further calculations we shall be limited by
consideration only linear terms in the $\varepsilon _ {(i) (j)} $.

Taking into account  that
$u\bar {^i} = (0,0,0, c)$ and $g_ {\bar {i} \bar {j}} (0) =\eta
_ {(i) (j)}$, we receive the following expressions for the system of
Mathisson---Papapetrou
equations in the rotating Fermi coordinates  (\ref {4}):
\begin {equation}
\mu W_ {(\alpha)} =-ce^ {(\mu) (\nu) (\tau)} R_ {(4) (\nu) (\alpha) (\mu)} ^ {}
\bar{S} _ {(\tau)} + ZeF_ {(\alpha) (4)}, \label {11} \end {equation}
\begin {equation} \frac d {d\tau} \bar{S} _ {(\alpha)} =e_ {(\alpha)
(\kappa) (\tau)} \omega ^ {(\tau)} \bar{S} ^ {(\kappa)},
\quad
\bar{S} _ {(4)} =0, \label {12} \end {equation}
where all quantities are taken  along the  nucleus world line ($X^ {(\alpha)}
=0$).

\section {Calculation of Electromagnetic Field of Electrons Inside the Atom.}

\ \ \ \ To find  how  the electromagnetic field of  electrons influences on the
atom motion, it is necessary, first, solve the Maxwell equations  in the Fermi
coordinates  for the electromagnetic field, created by electrons in the
arbitrary  point $\zeta ^ {(\alpha)}$. Second, calculate the expectation value
of that field, using the assumption, that the atom is in a quasistationary
quantum state.

Solving Maxwell equations \begin {equation}
\left [\sqrt {-\det (g_ {ij})} g^
{mk} g^ {nl} (A_ {l, k} -A_ {k, l}) \right] _ {, n} =0, \label {3.1}
\end{equation}
where $A_k= (A_ {(\alpha)}, A_4=-\varphi) $ is the electromagnetic
potentials, we will interest only quasistationary solutions, i.e. we shall
search  the field of  charges $e$, resting in the points $X_A^ {(\alpha)} $
($A=1, 2,..., Z$). Avoiding in this case the consideration of  the motion of
electrons we neglect the  effects of interaction of the magnetic field of
electrons with the nucleus magnetic momentum.  These effects are small enough
because of the  small  velocity  of  electrons  within  the  atom
($v \ll c$).

Writing down the Maxwell equations  in  the linear approximation in the
$\varepsilon _ {(i) (j)} $ and neglecting the terms, containing the derivative
on time of the potentials, we will have the following expression for electromagnetic
potentials of the system of the rest charges in some arbitrary point $\zeta ^
{(\alpha)}$:
$$
     \varphi   (\zeta ^{(\alpha )})=\sum\limits_{A=1}^{Z}
     \frac{e}{R_A}\left[1+\frac{1}{2c^2}W_{(\beta)}R_A^{(\beta)} +
     \frac{1}{6}R_{(4)(\mu)(4)(\nu )}\zeta^{(\mu )}\left(2\zeta^{(\nu)}-
X^{(\nu)}\right)+\right.
$$
\begin {equation} \label{3.3}
\left.
 +\frac{1}{6R_{A}^2}R_{(\beta)(\tau)(\nu)(\mu)}\zeta^{(\beta)}
\zeta^{(\nu)}X_A^{(\tau)}X_A^{(\mu)}+O(\rho _A^3)\right] ,
\end{equation}
\begin{eqnarray} \label{3.3a}
     A^{(\sigma)}(\zeta^{(\alpha)}) = \sum\limits_{A=1}^{Z}
     \frac{e}{R_A}\left[-\frac{1}{c}e^{(\sigma)}\,_{(\tau)(\nu)}
     X^{(\tau)}\omega^{(\nu)} -
     \frac{1}{6}R^{(\sigma)}\,_{(\nu)(4)(\tau)}\zeta^{(\nu)}
     \zeta^{(\tau)} + \right.\nonumber\\
     \left.
     +\frac{1}{2}R^{(\sigma)}\,_{(4)(\nu)(\tau)}\zeta^{(\nu)}
     X^{(\tau)}  +
     \frac{1}{2}R^{(\sigma)}\,_{(\tau)(\nu)(4)}X^{(\nu)}
     X^{(\tau)}+O(\rho _A^3) \right] ,
\end{eqnarray}
   where
$R_{A}^{(\alpha)}=(\zeta^{(\alpha)} - X_A^{(\alpha)})$,
$R_{A}=\sqrt{R_A^{(\alpha)}{R}_{A(\alpha)}}$,
$\rho _A=\sqrt{X_{A(\alpha )}X_A^{(\alpha )}}$.

Thus, we also assume, that
inside of our system there is no substance or other strong fields, resulting to
the additional curvature of space time, i.e. the Ricci tensor  inside of atom
is equal to zero ($R_ {i j} =0$). Otherwise, the presence of the additional
interaction would generate the effects, which exceed in  magnitude the
influence of the gravitational field on the atom motion.

Then,  in comoving  rotating Fermi coordinates the components of the tensor
of the average electromagnetic field will be equal to:
\begin {equation}
F_{(\alpha)(4)}=
    \left<{\hat{F}}_{(\alpha)(4)}\right> ,
\label{gor1}
 \end {equation}
where
$$
\hat{F} _ {(\alpha) (4)}\equiv
 -\left (\frac {\partial \hat{\varphi} (\zeta ^{(\kappa)})}
 {\partial \zeta ^ {(\alpha)} } \right) _ {\zeta ^ {(\nu)} =0}-
\frac{1}{c}\left( \frac {\partial\hat{A}_{(\alpha)}}{\partial\tau}
     \right)_{\zeta ^{(\nu)}=0} \approx
$$
\begin {equation}
    \approx -\sum\limits_ {A=1} ^Z\left (\frac {e\hat{X}_ {A (\alpha)}} {\hat{\rho} _A^3}
 +\frac {e}{\hat{\rho} _A} \left (\frac {1}{3}R_ {(\alpha) (4) (4) (\nu)} \hat{X}_A^ {(\nu)}
+ \frac {1}{2c^2} W_ {(\alpha)} \right) \right) .
\label {3.4} \end {equation}
Here, $\hat{X}_ {A (\alpha)}$ is the position operator of $A-$th electron;
$ \hat{\rho} _A\equiv\sqrt{\hat{X}_{A(\alpha )}\hat{X}_A^{(\alpha )}}$ and
the operators $\hat{\varphi}$ and $\hat{A}$ are obtained from  (\ref{3.3}) and  (\ref{3.3a})
by changing $X_{A}^{(\alpha)}\rightarrow \hat{X}_ {A}^ {(\alpha)}$,
$\rho_{A}\rightarrow \hat{\rho} _A$ and so on.

In order to determine the expectation value  (\ref{gor1}) of the electromagnetic
field tensor we will use the quasirelativistic two-component representation of
the covariant Dirac equation.
The detailed explanation  of this representation, and  the  consistent
formulation of the quantum mechanics in the arbitrary reference frames and in
the external gravitational fields can be found in the book \cite {gorb}.

According to this approach \cite {gorb}, the covariant Dirac equation can
be interpreted in general case as a special coordinate representation of the
traditional quantum-mechanical equation of motion in Hilbert space. Thus, the
problem about solving of the Dirac equation in the curved space-time can be
lead to  the similar problem of the traditional quantum mechanics in the flat
space - time, and the properties of the reference frame and the curved
space-time can be taken into account in the Hamiltonian as the additional
potentials.

In particular, in quasirelativistic approximation our system  can
be described by a two-component Pauli-type equation
\begin {equation}
i\hbar \frac{\partial \psi }{\partial \tau }=\hat{H}\psi
\end{equation}
(with
\begin {equation}
\int\limits_{f} \psi ^{+}\psi d^3\,X=1,
\end{equation}
and $d^3X=dX^{(1)}dX^{(2)}dX^{(3)}$) in the reference frame of single
observer.
The complete expression for the  two-component Hamiltonian  $\hat{H}$ (for one-electron atom)
can be found in \cite {gorb}.  The analysis of
 the complete operator $\hat{H}$ shows, that the influence
of external gravitational field will always be more significant than effects
caused by  the non-Coulomb terms of the nuclear potentials\footnote {The electromagnetic potentials
of the nucleus can be obtained from (\ref {3.3}) and (\ref {3.3a})
by the following substitution:
$e\rightarrow Ze$, $\zeta_ {A} ^ {\alpha} \leftrightarrow X_ {A} ^ {(\alpha)} $, and
 $\zeta_ {A} ^ {\alpha} \rightarrow 0$.}.

Therefore, in a good approximation the Hamiltonian of the system of $Z$ electrons
is given by:
\begin {equation} \hat{H}=\hat{H}_{0} +\hat{H}_{1}, \label {4.1}
\end{equation}
where
\begin {equation}
\hat{H}_{0}=\sum\limits_{A}\left(
\frac{\hat{P}_A^2}{2m}-\frac{Ze^2}{\hat{\rho} _A}\right)        +\frac
{1}{2}\sum\limits_{ A\neq  B}\frac{e^2}{\hat{\rho} _{AB}}
\label {4.2} \end {equation}
 is the  Hamiltonian  of
the  multielectron  atom  in  the  flat  space-time;
 $$   \hat{H}_{1}
  =\sum\limits_  {A}\left[ mc^2\hat{\theta}  _A    +
\frac{c}{2}\left(  e^  {(\sigma)  (\alpha)  (\tau)}
\hat{\theta} _ {A (\sigma), (\alpha)} \hat{S}_ {A (\tau)}
-\left\{\hat{\theta}  _A^{(\beta)}, {\hat{P}}_ {A (\beta)}
\right\} \right)-\right.
$$
\begin {equation}
-\left.\omega ^ {(\beta)} \left (\hat{L}_ {A
(\beta)} + \hat{S}_ {A (\beta)} \right) + mW_ {(\beta)} \hat{X}_A^{(\beta)} \right]
\label {4.3} \end {equation}
is  the Hamiltonian, describing the
interaction of the electrons with the external gravitational field.
Here we introduced the following notations:
$\hat{X}_A^{(\beta)}=X_A^{(\beta)}$,
 $\hat{P}_ {A (\alpha)} =\frac \hbar i\partial /\partial\hat{X}_A^ {(\alpha)} $,
$\hat{L}_ {A (\alpha)} =e_ {(\alpha) (\beta) (\tau)} \hat{X}_A^ {(\beta)}\hat{P}_A^ {(\tau)} $
and $\hat{S}_ {A(\alpha)} =\frac \hbar 2 \sigma _ {(\alpha)} $
 are the operators of the position, momentum, angular momentum and spin for the $A$-th  electron;
$A=1, 2,..., Z$;
$\sigma _{(\alpha)} $  are the standard Pauli matrices;
$\hat{\theta} _ {A (\alpha)} =\frac {2}{3}R_ {(\alpha) (\mu) (\nu) (4)}\hat{X}_A^ {(\mu)}\hat{X}_A^
{(\nu)} $;
$\hat{\rho} _ {AB} =\sqrt {(\hat{X}_ {A (\alpha)} -\hat{X}_ {B (\alpha)})
 (\hat{X}_A^ {(\alpha)} -\hat{X}_B^{(\alpha)})} $.

From (\ref {3.4}) by using the symmetry of the
non-disturbed stationary state of the atom, we have
\begin {eqnarray} \label {gor3}
    \left < \hat {F} _ {(\alpha) (4)} \right > _ {0}
     = - \frac {e} {2c^ {2}} W_ {(\alpha)}
         \left < \sum\limits_ {A}
     \frac {1} {\hat {\rho} _ {A}} \right > _ {0} \;,
\end {eqnarray}
    where notation $\left\langle ... \right\rangle_0$ means that the expectations values are
calculated using eigenvectors of non-disturbed Hamiltonian $\hat{H}_{0}$
 (\ref {4.2}). If one neglects ``deformation'' of the atom by external gravitational field,
then the electromagnetic field
 $F _ {(\alpha) (4)}$
 (see (\ref {gor1}), (\ref {11})) will affect only renormalization of
 the nucleus mass $\mu \rightarrow \mu + \triangle \mu$, where
$ \triangle\mu=
\frac {Ze^2} {2c^2} \left\langle
\sum\limits_ {A} \frac {1} {\hat {\rho} _ {A}} \right\rangle _ {0}$.
Because of  $ \triangle\mu/\mu\ll 1$, we will not  hereinafter consider this renormalization.

Using  perturbation theory, it is easy to show, that the required
expectation  value  of  the  electromagnetic  field  tensor  from
equation (\ref{3.4})
can be presented as \cite{gorb1}:
\begin  {equation}
F_{(\alpha) (4)}= \left\langle \hat {F} _  {(\alpha)  (4)}
\right\rangle =  \left\langle  \sum\limits_  {A}
\left (-\frac {e\hat{X}_ {A (\alpha)}} {\hat{\rho} _A^3} \right) \right\rangle
+  O(\varepsilon  _  {(\alpha)  (\beta)}  ^2),  \label  {4.4}
\end {equation}
where $O(\varepsilon  _{(\alpha)(\beta)}  ^2)$  denote
the terms quadratic in $\varepsilon _ {(i)(j)}$.

Direct calculation of quantity
$\left\langle \sum\limits_ {A} \left (-\frac  {e\hat{X}_ {A (\alpha)}}
 {\hat{\rho} _A^3} \right) \right\rangle$
is connected with
mathematical difficulties in the framework of the perturbation theory.
Therefore we shall use the following property of the hermitian operators.

If observable $L$ is described by the hermitian operator $\hat{L}$, which does not depend
on time explicitly, the expectation value of the operator of  evolution $ <
\stackrel {\circ} {L} > $ of this  observable $L$ in the stationary state (for
which the state vector is an eigenvector of the Hamiltonian) will be equal to
zero: \begin {equation} \frac d {dt} <\hat{L}> = < \stackrel {\circ} {L} > =0,
\label{4.5} \end {equation}
where
\begin {equation}
\stackrel {\circ} {L} \equiv \left (\frac {\partial\hat{L}} {\partial
\tau} \right) _ {\exp l} + \frac i {\hbar} \left [\hat{H}, \hat{L}\right] .
\label {4.6} \end {equation}

 In  general case the Hamilton operator of the atom in an external gravitational field
always explicitly depends on proper time $\tau$.
But, we can decompose  the Hamiltonian $H$ of our quantum system in a vicinity of
an arbitrary  moment of time $\tau _0$:
$$
H (\tau) =H (\tau _0) + \left (\frac {\partial H} {\partial \tau} \right) _ {\tau =\tau _0} \left (\tau -\tau _0\right) +...
$$
and  calculate the probabilities of transitions, caused by the term
 $\left (\frac {\partial H} {\partial \tau} \right) _ {\tau =\tau _0} \left (\tau -\tau _0\right) $
 in the  operator $H$. If we will find, that
during the time $\left (\tau -\tau _0\right) \sim \Delta \tau $
(where $\Delta\tau $ is the characteristic atomic time)
these probabilities are much less than 1,
we will consider the eigenvector of the operator $H$  as the quasistationary state of the atom.
Numerical  estimations show, that this quasistationary conditions for atom states are satisfied
 for a wide class of gravitational fields (such as the external gravitational field of the macroscopic black holes).
In our case, the terms included in the Hamiltonian $ \hat{H}_{1}$,
such as $\hat{\theta}$ and $\hat{\theta} _ {(\alpha)}$, are very slowly change with time.
Therefore, even in case of the ultrarelativistic motion of the atom, when the
atom passes a short distances during a
characteristic atomic time, the gravitational field  can be consider as
homogeneous with a high precision (except of the  case of short-wave
gravitational radiation).

 We can introduce
 the operators of 3-velocity $\stackrel{\circ}{X} _ {A(\alpha)} $:
$$
\stackrel{\circ}{X}_{A(\alpha )}\equiv\left( \frac{\partial \hat{X}_{A(\alpha )}}{\partial
\tau }\right) _{\exp l}+\frac i{\hbar} \left[ \hat{H}, \hat{X}_{A(\alpha )}\right]\simeq
$$
\begin{equation}
\simeq \frac {1}{m}\hat{P}_{A(\alpha )}^{}-\frac 23cR_{(\alpha )(\mu )(\nu )(4)}
\hat{X}_A^{(\mu )}\hat{X}_A^{(\nu )}-
e_{(\alpha )(\beta )(\tau )}^{}\omega ^{(\beta )}\hat{X}_A^{(\tau )},
\label {x} 
\end{equation}
and 3-acceleration $\stackrel{\circ\circ}{X} _ {A (\alpha)} $:
$$
\stackrel{\circ\circ}{X}_{A(\alpha )}\equiv\left( \frac{\partial \stackrel{\circ}{X}
_{A(\alpha )}}{\partial \tau }\right) _{\exp l}+\frac i {\hbar} \left[ \hat{H},
\stackrel{\circ}{X}_{A(\alpha )}\right]\simeq
$$
$$
\simeq \frac {1}{m}\left( \frac{Ze^2\hat{X}_{A(\alpha )}}{\hat{\rho} _A^3}\right)+\frac
1m\sum\limits_{B, B\neq A}\left( \frac{e^2\left( \hat{X}_{A(\alpha )}-\hat{X}_{B(\alpha
)}\right) }{2\hat{\rho} _{AB}^3}\right) -c^2\hat{\theta} _{A,(\alpha )}-
$$
$$
-\frac c{2m}e^{(\sigma )(\beta )(\tau )}\hat{\theta}_{A(\sigma ),(\beta ),(\alpha
)}\hat{S}_{A(\tau )}-\frac{ci}{2m\hbar }\left[ \left\{ \hat{\theta} _A^{(\beta
)}, \hat{P}_{A(\beta )}\right\} , \hat{P}_{A(\alpha )}\right] +
$$
\begin{equation}
+\frac 1m\omega ^{(\beta )}e_{(\beta )(\alpha )(\tau )} \hat{P}_A^{(\tau
)}-W_{(\alpha )}
\label {xx} 
\end{equation}

By direct calculation it is possible to show, that, if the atom is
in a quasistationary state, the operators of 3-velocity $\stackrel{\circ}{X} _ {A
(\alpha)} $ and 3-acceleration $\stackrel{\circ\circ}{X} _ {A (\alpha)} $ for each
electron satisfy to the following equations:
\begin {equation} \left\langle
\stackrel{\circ}{X} _ {A (\alpha)}  \right\rangle  =0,
\quad
\left\langle \stackrel{\circ\circ}{X}  _  {A  (\alpha)}
\right\rangle =0. \label{4.7} \end {equation}

Taking into account the obvious explicit time-independence  of the operators
$\hat{X}_{A} ^{(\alpha)} $ and $\hat{P}_ {A} ^ {(\alpha)} $ and the fact, that they satisfy to
following commutation  relations:
\begin {equation} \left [\hat{P}_A^ {(\alpha)},
\hat{X}_B^ {(\beta)} \right] =\frac \hbar i\delta ^ {(\alpha) (\beta)} \delta _ {AB},
\label {4.8} \end {equation} \begin {equation} \left [\hat{X}_A^ {(\alpha)}, \hat{X}_B^
{(\beta)} \right] =0,
\quad \left [\hat{P}_A^ {(\alpha)},\hat{P}_B^ {(\beta)}
\right] =0, \label {4.9} \end {equation}
we receive from system of the
equations (\ref{4.7}) the following expression:
$$
\left\langle \sum\limits_{A}\left( -\frac{Ze^2\hat{X}_{A(\alpha )}}{\hat{\rho} _A^3}
\right) \right\rangle \cong -ZmW^{(\alpha )}-
$$
\begin{equation}
-\left\langle \sum\limits_{A}\left[
mc^2\hat{\theta} _{A,(\alpha )}
-ce^{(\sigma )(\delta )(\tau )}R_{(\alpha )(\delta )(4)(\sigma)}\left( \hat{S}_{A(\tau )}+\hat{L}_{A(\tau )}\right)
\right] \right\rangle ,
\label {4.10}
\end{equation}
where we used, that  ``gravitational
potentials''  $\hat{\theta}$ and $\hat{\theta} _ {(\alpha)} $ very slowly vary with time:
\begin {equation}
m\left\langle \stackrel{\circ} {\theta} _ {(\kappa)} \right\rangle
\cong R_ {(4)(\mu) (\kappa) (\nu)} \left\langle \left\{\hat{X}^ {((\mu)}, \hat{P}^ {(\nu))} \right\}
\right\rangle \cong 0.
\label {4.11} \end {equation}

Because of we neglect of quadratic terms in $\varepsilon _ {(i) (j)}$, we can
change from $\left\langle... \right\rangle$ to $\left\langle ... \right\rangle
_0$, where notation $\left\langle... \right\rangle _0$ means that the
expectation value is calculated in relation to eigenvectors of non-disturbed
operator (\ref {4.2}). Then  we receive from the system (\ref {11}) (\ref {12})
the following expression for the equation of  motion of the multielectron atom
in an external gravitational field in the comoving reference frame:
$$
\left (\mu + Zm\right) W_ {(\alpha)} =-ce^ {(\sigma) (\nu) (\tau)}
 R_ {(4)(\nu) (\alpha) (\sigma)} ^ {} \left [\bar{S} _ {(\tau)} + \right.
$$
\begin {equation} \left.
+ \left\langle \sum\limits_ {A}\left (\hat{S}_ {A (\tau)} + \hat{L}_ {A (\tau)}
\right) \right\rangle _0\right]
-mc^2\left\langle \sum\limits_ {A} \hat{\theta} _ {A, (\alpha)} \right\rangle_0.
\label {4.14} \end {equation}

 The second term in the right part of the received equation (\ref {4.14})
$$
 -mc^2\left\langle
\sum\limits_ {A}\hat{\theta} _ {A, (\alpha)} \right\rangle _0=
\frac{1}{4}mc^2 \left[R_{(4)(\mu)(4)(\alpha)(\tau)}+\right.$$
\begin {equation}
\left.+R_{(4)(\tau)(4)(\mu)(\alpha )}+
R_{(4)(\alpha )(4)(\tau)(\mu)}\right]
\left< \sum\limits_{A}\hat{X}_{A}^{(\mu)}\hat{X}_{A}^{(\tau)}\right>_{0}
\end {equation}
describes the interaction of an external gravitational field with
the average quadruple moment of atom.
From estimations  made above, this term becomes actually small for atomic
dimensions, if  we assume, that the full spin of the atom (which
consists  of the spins of the nucleus and electrons, and also of  the
electronic orbital moment) is not equal to zero. The calculations of the expectation
values of the operator
$\sum\limits_{A}\left(\stackrel{\circ}{L}_{A (\alpha)}+
    \stackrel{\circ}{S}_{A (\alpha)}\right)$ describing variation of the electron
angular momentum, by means of (\ref {3}) lead us to
$$
     \frac{d}{d\tau}\left[\bar{S}_{(\alpha )}
     + \left<\sum\limits_{A}( \hat{L}_{A(\alpha )}+ \hat{S}_{A(\alpha )})
     \right>\right]
     \cong
$$
\begin{eqnarray} \label{gor4}
 \cong     e_{(\alpha )}\,^{(\nu)(\kappa)}\omega_{(\kappa)}
\left[\bar{S}_{(\nu )}
     + \left<\sum\limits_{A}(\hat{L}_{A(\nu )}+ \hat{S}_{A(\nu )})
     \right>_{0}\right] .
\end{eqnarray}

Finally, the equation of  motion of the multielectron atom it is possible to
write down as:
$$ \left (\mu + Zm\right) W_ {(\alpha)} =-ce^ {(\sigma) (\nu) (\tau)} R_ {(4)
(\nu) (\alpha) (\sigma)} ^ {} \left [\bar{S} _ {(\tau)} + \right. $$
\begin {equation} \left.
+ \left\langle \sum\limits_ {A} \left ( \hat{S}_ {A (\tau)} + \hat{L}_ {A (\tau)}
\right) \right\rangle _0\right].
\label {4.15} \end {equation}

In this way, we have demonstrated, using the some  physically reasonable
assumptions, that the motion of multielectron atom in an external gravitational
field in a good approximation can be described by the system of  classical
Mathisson---Papapetrou equations
\begin {equation}
\left(\mu + Zm\right) \frac  {Du^i}{D\tau}=\frac 1 {2c} R^{i}\,_ {jkl}u^j\varepsilon ^ {klmn} u_n J_m,
 \label {4.16} \end {equation}
 \begin {equation}
\frac {DJ^i}{D\tau}=
      \frac{1}{c^2}u^iJ_n\frac  {Du^i}{D\tau},\quad
J_m u^m=0,
\label{4.17} \end {equation}
if we put as a classical angular momentum of
atom the expectation value of the operator $J^k$ of the total angular momentum of the
system. In comoving  rotating Fermi coordinates the quantity $J^k$
 has the following components:
\begin{equation}\label {4.18}
J _{(\alpha )}=\bar{S}_{(\alpha)}+\left<
\sum\limits_{A}(\hat{L}_{A(\alpha)}+\hat{S} _{A(\alpha )}) \right>_0,
\quad
J_{(4)}=0.
\end{equation}

We note that if we also consider the influence of any external electromagnetic field $\; ^{(0)}{F^{ij}}$ on the atom motion,
we have to add in the equation  (\ref {4.16}) the following terms:
\begin{equation}\label {4.19}
\frac {e}{c}(Z+1)\; ^{(0)}{F^{ij}}u_j+\frac {1}{c}\; ^{(0)}F^{i}\,_{m;n}
\varepsilon^{mnkl}M_{k}u_{l},
\end{equation}
where $M_{k}$ is the 4-vector of the atoms full magnetic moment, which in comoving reference
frame has the  the following components:
\begin{equation}\label {4.20}
M_{(\alpha )}=\frac e{2mc}\left\langle {\sum\limits_{A}}
\left(\hat{\cal L}_{(\alpha )}+2\hat{S}_{(\alpha )}\right) \right\rangle _0+\frac{Zeg}{2\mu c}
\bar{S}_{(\alpha )},
\quad
M_{(4)}=0.
\end{equation}

As it is visible from the received equation (\ref {4.15}), the trajectory of
the atom will depend on the  total angular momentum of the  atom, which
includes spins of the  nucleus and electrons, and also orbital moment of
electrons in atom.
Hence, the motion of the atom in many  respects  depends  on  his
quantum state.  So, for example, the atoms, being in  states  with
different mutual orientations of spins of the nuclear  electrons,
will move in general case along the  different
trajectories in an external  gravitational  field.  Besides,  the
atoms, being in excited states  with  long  life  time,  will  be
separated by an external gravitational field  from  atoms,  which
being in nonexcited states.
Thus, the ultrarelativistic  motion  of
atoms in the external gravitational fields can lead to a  set  of
generally-relativistic  effects,  which can represent   interest    for
astrophysics and allow us to check  up  the  predictions  of  the
general relativity on the quantum level.


\begin {thebibliography} {99}

\bibitem {dalgar} Dalgarno A. In {\em Rydberg states of Atoms and Molecules},
Cambridge, 1983.

\bibitem {pino1} Pinto F. {\em Phys. Rev. Lett.}, {\bf 70}, 3839 (1993).

\bibitem {pino2} Pinto F. {\em Gen. Rel. Grav.}, {\bf 27}, 9 (1995).

\bibitem {parker1} Parker L., Vollick D.  {\em  Phys.Rev.},  {\bf
D56},  2113-2117 (1997).

\bibitem {parker2} Parker L. {\em Phys.Rev.}, {\bf  D22}, 1922-1934 (1980).


\bibitem {parker3} Parker L. {\em Phys.Rev. Lett.}, {\bf 44}, 1559 (1980).
\bibitem {audr}Audretsch J. {\em J. Phys. A: Math. Gen}, {\bf 14}, 411 (1981).

\bibitem {gorb}  Gorbatsievich A. K. {\em  Quantum Mechanics in General  Theory
of Relativity.}, Minsk, University Press 1985 (in Russian).

\bibitem {gorb1} Gorbatsievich A. K. {\em Acta Phys. Polon.}, {\bf B17}, 111 (1986).

\bibitem {gorb2} Gorbatsievich A. K. {\em Acta Phys. Polon.}, {\bf B16}, 21 (1985).

\bibitem {fish}
Fischbach E, Freeman B.S., Weu-Kwei Cheng., {\em Phys.Rev.}, {\bf D23}, 2157 (1980).

\bibitem {mathis} Mathisson M. {\em Acta Phys. Polon.}, {\bf 6},  163-169 (1937).

\bibitem {papap} Papapetrou A. {\em Proc. Roy. Soc. (London)}, {\bf A209}, 248- 268
(1951).

\bibitem {dix} Dixon W.G. {\em Nuovo Cimento}, {\bf 34}, 317 (1964).

\bibitem {piran} Pirani F. A. E. {\em Acta Phys. Polon.}, {\bf 15}, 389-405 (1956).

\bibitem {tulcz}Tulczyjew W. {\em Acta Phys. Polon.}, {\bf 18}, 37-55 (1959).

\bibitem {sing}Singe J. K. {\em Relativity: The General Theory}, Amserdam,
1960.

\bibitem {mizn} Misner C. W., Thorne K. S., Wheeler J. A. {\em Gravitation}, San Fransisco, 1973.

\bibitem {ni} Ni W.-T., Zimmermann M. {\em Phys. Rev.}, {\bf 17}, N. 6, p. 1473, (1978).

\end{thebibliography}
\end {document}